\documentstyle[preprint,aps,epsfig]{revtex}
\begin{document}

%\titlepage
\draft
\tighten
\preprint{
\vbox{
\hbox{ADP-99-53/T389}
\hbox{IU/NTC 00-02}
}}

\title{Lambda Polarization in Polarized Proton-Proton Collisions 
at RHIC}

\normalsize
\author{C. Boros, J.T. Londergan$^1$ and  A.W. Thomas}
\address {Department of Physics and Mathematical Physics,
                and Special Research Center for the
                Subatomic Structure of Matter,
                University of Adelaide,
                Adelaide 5005, Australia}

\address{$^1$ Department of Physics and Nuclear
            Theory Center, Indiana University,
            Bloomington, IN 47408, USA}

\date{\today}
\maketitle

\begin{abstract}
We discuss Lambda polarization in 
semi-inclusive proton-proton collisions,  
with one of the protons longitudinally polarized. 
The hyperfine interaction responsible for the $\Delta$-$N$ and 
$\Sigma$-$\Lambda$ mass  splittings gives rise to flavor asymmetric 
fragmentation functions and to sizable 
polarized non-strange fragmentation functions. 
We predict large positive Lambda 
polarization in polarized proton-proton collisions 
at large rapidities of the produced  Lambda, while other models,  
based on $SU(3)$ flavor symmetric fragmentation functions, predict 
zero or negative Lambda polarization. 
The effect of $\Sigma^0$ and $\Sigma^*$ decays 
is also discussed. 
Forthcoming experiments 
at RHIC will be able to differentiate between these 
predictions. 
\end{abstract}

\newpage

\section{Introduction} 
Measurements of the polarization dependent 
structure function, $g_1$, in deep inelastic scattering \cite{g1}  
have inspired considerable experimental and theoretical effort to  
understand the spin structure of baryons. While 
most of these studies concern 
the spin structure of the nucleons, it has  become clear that 
similar measurements involving  other baryons would provide 
helpful, complementary information 
\cite{Burk,Nzar,Ellis,deFlorian,Boros,Ma,Lipkin,BorosLiang}.    
The Lambda baryon plays a special role in this respect. 
It is an ideal testing ground for spin studies 
since it has a rather simple  
spin structure in the naive quark parton model. 
Furthermore, its self-analyzing decay makes polarization 
measurements experimentally feasible. 

Forthcoming  
experiments at RHIC could measure the polarization of 
Lambda hyperons produced in proton-proton collisions 
with one of the protons longitudinally  polarized, 
$p^\uparrow p \rightarrow \Lambda^\uparrow X$. 
The polarization dependent fragmentation function 
of quarks and gluons into Lambda hyperons can be extracted 
from such experiments.  
These fragmentation functions contain information on how the spin of the 
polarized quarks and gluons  is transferred 
to the final state Lambda. The advantage of proton proton 
collisions, as opposed to $e^+e^-$ annihilation, 
where $\Lambda$  production and polarization is dominated by strange quark 
fragmentation,  
is that Lambdas at large positive rapidity are 
mainly fragmentation 
products of up and down valence quarks of the polarized  
projectile.  
Thus, the important question, 
intimately related to our understanding of the spin structure 
of baryons, of whether polarized up and down quarks 
can transfer polarization to the Lambda  
can be tested at RHIC \cite{deFlorian}.

In a previous publication, we have shown that 
the hyperfine interaction, responsible for the
$\Delta$-$N$ and  $\Sigma^0$-$\Lambda$ mass splittings   
leads to non-zero polarized non-strange quark  
fragmentation functions \cite{Boros2}. These non-zero polarized 
up and down quark fragmentation functions give rise to 
sizeable {\it positive} $\Lambda$ polarization in experiments 
where the strange 
quark fragmentation is suppressed. 
On the other hand, predictions based either on 
the naive quark model or on $SU(3)$ flavor symmetry  
predict {\it zero} or {\it negative} Lambda polarization \cite{Burk}. 

In section II, we briefly discuss fragmentation functions and 
show how the hyperfine interaction leads to polarized non-strange 
fragmentation functions. We fix the parameters of the model by 
fitting the data on $\Lambda$ production in 
$e^+e^-$ annihilation. In section III, we discuss 
$\Lambda$ production in pp collisons at RHIC energies.  
We point out that the production of $\Lambda$'s at high 
rapidities is dominated by the fragmentation of  
valence up and down quarks of the polarized 
projectile, and is ideally suited to test whether non-strange 
quarks transfer their polarization to the final state $\Lambda$.  
We predict significant positive $\Lambda$-polarization 
at large rapidities of the produced $\Lambda$.

\section{Fragmentation functions} 

Fragmentation functions can be defined 
as light-cone Fourier transforms of matrix elements of quark operators
\cite{Collins,JaffeJi}
\begin{equation}
 \frac{1}{z} D_{q\Lambda}^{\,\Gamma}(z)
   = \frac{1}{4} \sum_n
   \int \frac{d\xi^-}{2\pi}
      e^{-iP^+ \xi^-/z}
         \mbox{Tr} \{ \Gamma\, \langle 0|\psi (0) |
 \Lambda (PS);n (p_n)  \rangle
        \langle \Lambda (PS); n (p_n)
| \overline{\psi} (\xi^-) | 0\rangle \},
\label{frag1}
\end{equation} 
where, $\Gamma$ is the appropriate Dirac matrix; 
$P$ and $p_n$ refer to the 
momentum  of the produced $\Lambda$  and of the intermediate system $n$; 
$S$ is the spin of the Lambda and the plus projections of the 
momenta are defined by  $P^+\equiv \frac{1}{\sqrt{2}}(P^0+P^3)$. 
$z$ is the plus momentum fraction of the quark carried by the 
produced $\Lambda$. 

Translating the matrix elements, using the integral
representation of the delta function and projecting out the
light-cone plus and  helicity $\pm$ components 
we obtain
\begin{equation}
 \frac{1}{z} D^{\pm}_{q\Lambda} (z)
    =  \frac{1}{2\sqrt{2}}\sum_n
    \delta [(1/z-1)P^+-p_n^+]
        | \langle 0|\psi_{+}^{\pm} (0)
|\Lambda (PS_\parallel); n (p_n)   \rangle |^2 .
\label{frag2}
\end{equation}
Here, $\psi^\pm_+ = \frac{1}{2}\gamma_-\gamma_+ 
\frac{1}{2}(1\pm \gamma_5)\psi$,  and we have defined 
$\gamma_\pm = \frac{1}{\sqrt{2}} (\gamma_0 \pm \gamma_3)$.  
The fragmentation function of an anti-quark
into a $\Lambda$ is given by Eq. (\ref{frag2}), with
$\psi_+$ replaced  by $\psi_+^\dagger$: 
\begin{equation}
 \frac{1}{z} D^{\pm }_{\bar q\Lambda} (z)
    =  \frac{1}{2\sqrt{2}} \sum_n
    \delta [(1/z-1)P^+-p_n^+]
        | \langle 0|\psi_{+}^{\dagger\pm} (0)
|\Lambda (PS_\parallel); n (p_n)   \rangle |^2 .
\label{frag3}
\end{equation}
$D_{q\Lambda}^{\pm}$ can be interpreted as
the probability that 
a quark with positive/negative helicity fragments into a $\Lambda$ 
with positive helicity and similarly for antiquarks.

The operator $\psi_+$ ($\psi_+^\dagger$) either destroys a quark 
(an antiquark) or it creates an antiquark (quark) 
when acting on the $\Lambda$ on the right hand side in the matrix elements. 
Thus, whereas, in the case of quark fragmentation, the intermediate state 
can be either an anti-diquark state, $\bar q\bar q$, or a 
four-quark-antiquark state, $q\bar q\bar q\bar q$, 
in the case of antiquark fragmentation, only four-antiquark states, 
$\bar q\bar q\bar q\bar q$, are possible assuming that there are no 
antiquarks in the $\Lambda$. (Production of $\Lambda$'s through coupling 
to higher Fock states of the $\Lambda$ is more complicated and 
involves higher number of quarks in the intermediate states. 
As a result it would lead to  
contributions at lower $z$ values.)   
Thus, we have 
\begin{itemize} 
 \item (1a) $q\rightarrow qqq + \bar q\bar q = \Lambda + \bar q \bar q$ 
 \item (1b) $q\rightarrow qqq + q \bar q \bar q\bar q 
 = \Lambda + q \bar q \bar q\bar q$,  
\end{itemize} 
for the quark fragmentation  and 
\begin{itemize} 
\item  (2) $\bar q\rightarrow qqq + \bar q \bar q \bar q\bar q 
= \Lambda + \bar q \bar q \bar q\bar q$, 
\end{itemize}  
for the antiquark fragmentation. 

While, in case (1a), the initial fragmenting quark is contained 
in the produced Lambda, in case (1b) and (2), the Lambda 
is mainly produced by quarks created in the fragmentation 
process.  
Therefore, we not only expect that Lambdas produced through (1a) 
usually have  larger momenta than those produced 
through (1b) or (2) but also that Lambdas produced through 
(1a) are much more sensitive to the flavor spin quantum numbers 
of the fragmenting quark than those produced through 
(1b) and (2). In the following we assume  
that (1b) and (2) lead to approximately 
the same fragmentation functions. In this case, the difference, 
$D_{q\Lambda} - D_{\bar q\Lambda}$, 
responsible for leading particle production, 
is given by the fragmentation functions associated with process (1a).

Similar observations also follow  from energy-momentum 
conservation built in Eqs. (\ref{frag2}) and  (\ref{frag3}).    
The delta function  implies that
the  function, $D_q(z)/z$, peaks at \cite{Boros2}  
\begin{equation}
   z_{max} \approx \frac{M}{M +M_n}. 
\label{max}
\end{equation} 
Here, $M$ and $M_n$ are the mass of the produced
particle and the produced system, $n$, and
we work in the rest frame of the produced particle. 
We see that the location of  the maxima of the fragmentation function
depends on the mass of the system $n$.  While the high $z$
region is dominated by the fragmentation of a quark into the
final particle and a small mass system, large mass systems
contribute to the fragmentation  at lower $z$ values. 
The maxima of the fragmentation functions from process (1a) are 
given by the mass of the intermediate diquark state and that 
of the  the fragmentation functions from the processes 
(1b) and (2b) by the masses of intermediate four quark 
states. Thus, the contribution from process (1a) 
is harder than those from (1b) and (2).

Energy-momentum conservation also requires that 
the fragmentation functions are not flavor symmetric.  
While the assertion of  isospin symmetry, 
$D_{u\Lambda} =D_{d\Lambda}$, is well justified,  SU(3) 
flavor symmetry is broken not only by the strange quark mass but also 
by the hyperfine interaction. Let us discuss the fragmentation of a 
$u$ (or $d$) quark and that of an $s$ quark into a Lambda through 
process (1a). While the intermediate diquark state is always a scalar  
in the strange quark fragmentation, it can be either a 
vector or a  scalar diquark in the fragmentation of the non-strange 
quarks. The masses of the scalar and vector non-strange 
diquarks follow from 
the mass difference between the nucleon and the Delta \cite{Close}, while  
those of the scalar and vector diquark containing a strange quark 
can be deduced 
from the mass difference between $\Sigma$ and $\Lambda$ \cite{Alberg}.   
They are roughly $m_s \approx 650$
MeV and $m_v \approx 850$ MeV for the scalar and vector non-strange diquarks, 
and $m_s^\prime \approx 890$ MeV
and $m_v^\prime \approx 1010$ MeV for scalar and vector diquarks containing 
strange quarks, respectively \cite{Alberg,Boros2}.   
According to Eq. (\ref{max}), 
these numbers lead 
to soft up and down quark fragmentation functions 
and to hard  strange quark fragmentation functions. 

Energy-momentum conservation, together with the splitting of 
vector and scalar diquark masses,  
has the further important consequence 
that polarized non-strange quarks can transfer polarization to the 
final state Lambda. 
To see this  we note that the probabilities for 
the intermediate state to be  a scalar or vector 
diquark state in the fragmentation of an up or down quark 
with parallel or anti-parallel spin to the  spin of the Lambda 
can be obtained from the $SU(6)$ wave function of the $\Lambda$ 
\begin{eqnarray}
  \Lambda^\uparrow & = &
         \frac{1}{2\sqrt{3}}   [2 s^\uparrow  (ud)_{0,0}
+ \sqrt{2} d^\downarrow  (us)_{1,1}
 -d^\uparrow  (us)_{1,0} + d^\uparrow  (us)_{0,0} +\nonumber \\
  &&  -\sqrt{2} u^\downarrow   (ds)_{1,1}
 +u^\uparrow  (ds)_{1,0} - u^\uparrow    (ds)_{0,0} ]\,\,.
\end{eqnarray}
While the $u$ or $d$ quarks
with  spin  anti-parallel to the spin of the $\Lambda$ 
are always associated with a vector diquark,
$u$ and $d$ quarks with parallel spin 
have equal probabilities to be  
accompanied by a vector or scalar diquark.   
The fragmentation functions of non-strange quarks
with spin parallel to the $\Lambda$ spin are harder than
the corresponding fragmentation functions with anti-parallel spins. 
Thus, $\Delta D_{u\Lambda}$ is positive for large $z$ 
values and negative for small $z$. Their total contribution 
to polarized Lambda production might be zero or very small. 
Nevertheless, $\Delta D_{u\Lambda}$ and $\Delta D_{d\Lambda}$
can be sizable for large $z$ values, since both 
$D_{u\Lambda}$ and $\Delta D_{u\Lambda}$ are
dominated by the spin-zero component in the large $z$ limit. 
Furthermore, they will dominate polarized Lambda production whenever 
the production from strange quarks is suppressed. 

The matrix elements can be calculated 
using model wave functions  at the scale relevant to the
specific model and the resulting fragmentation functions can be evolved
to a higher scale to compare them to experiments. In a previous paper 
\cite{Boros2},   
we calculated the fragmentation functions in the MIT bag model and 
showed  that the resulting fragmentation functions  give 
a very reasonable description of the data in $e^+e^-$ annihilation. 
Since the mass of the intermediate states containing more than 
two quarks are not known we only calculate  the contributions of the 
diquark intermediate states in the bag model. The other contributions 
have been  determined by performing a global fit to the 
$e^+e^-$ data. For this,  we used the simple functional form 
\begin{equation}
      D_{\bar q\Lambda} (z)  =  N_{\bar q} z^\alpha (1-z)^\beta  
\label{para} 
\end{equation} 
to parameterize 
$D_{\bar q \Lambda} = D_{\bar u\Lambda} = D_{\bar d\Lambda} =...
D_{\bar b\Lambda}$ and also set  $D_{g\Lambda}=0$ 
at the initial scale, $\mu =0.25$ GeV. 

The fragmentation functions have to be evolved to the scale 
of the experiment, $\mu$. 
The evolution of the non-singlet fragmentation functions in LO 
is given by \cite{Owens,Uematsu}  
\begin{equation} 
  \frac{d}{d\ln \mu^2} [D_{q\Lambda} - D_{\bar q \Lambda} ](z,\mu^2)   
 = \int_z^1 \frac{dz}{z^\prime} P_{qq}(\frac{z}{z^\prime} )   
  [ D_{q\Lambda} - D_{\bar q \Lambda}](z^\prime ,\mu^2).   
\end{equation} 
The singlet evolution equations are  
\begin{eqnarray} 
  \frac{d}{d\ln \mu^2} \sum_q D_{q\Lambda} (z,\mu^2) 
& = & \int_z^1 \frac{dz^\prime}{z^\prime } 
 [ P_{qq}(\frac{z}{z^\prime}) \sum_q D_{q\Lambda} (z,\mu^2) + 
 2 n_f P_{gq} (\frac{z}{z^\prime }) D_{g\Lambda} (z^\prime ,\mu^2) 
]\nonumber \\ 
  \frac{d}{d\ln \mu^2} D_{g\Lambda} (z,\mu^2) 
& = & \int_z^1 \frac{dz^\prime}{z^\prime } 
 [ P_{qg}(\frac{z}{z^\prime}) \sum_q  
 D_{q\Lambda} (z,\mu^2) + 
 P_{gg} (\frac{z}{z^\prime }) D_{g\Lambda} (z^\prime ,\mu^2) ],  
\end{eqnarray} 
where the splitting functions are the same as those for 
the evolution of quark distributions.  
$n_f$ is the number of flavors.  
We used the evolution package of Ref. \cite{Kumano}  
suitably modified for the evolution 
of fragmentation functions (interchanging the off-diagonal elements in 
the singlet case).    

The results of the LO fit to $e^+e^-$ data 
\cite{SLD,Aleph,Delphi,L3,Opal,Lafferty}  
are shown in Fig.~1.   
The parameters of our fits are given in Table~I.   
The bag model calculations   
for $D_{q\Lambda}-D_{\bar q\Lambda}$ and 
$\Delta D_{s\Lambda}- \Delta D_{\bar s\Lambda}$ were parametrized 
using the  functional  form of Eq.~(\ref{para}).  
The fragmentation functions, $\Delta D_{u\Lambda}- \Delta D_{\bar u \Lambda}
=\Delta D_{d\Lambda}- \Delta D_{\bar d \Lambda}$, 
change sign at some value of $z$, hence we parametrized them using 
the form  
\begin{equation} 
      \Delta D_{q\Lambda}(z)   - \Delta D_{\bar q\Lambda}(z)   
=  N_{\bar q} z^\alpha (1-z)^\beta 
        (\gamma-z).    
\label{dpara} 
\end{equation}
These parameters are also given in Table~I.  
We also performed a fit using flavor symmetric fragmentation functions 
which we shall need for the discussion of Lambda production in 
pp collisions. The fit parameters are given in Table.~II. 
In Fig. 2, we show the calculated   
fragmentation functions at $Q^2=M_Z^2$.  
We note that 
our fragmentation functions describe  the 
asymmetry in leading and non-leading particle 
production, as well as  the lambda polarization measured in 
$e^+e^-$ annihilation 
at the $Z$ pole, very well --- as has been shown in Ref. \cite{Boros2}. 

\section{Polarized proton proton collision} 

Spin-dependent fragmentation of quarks can be 
studied in proton-proton collisions with one of the 
protons polarized \cite{deFlorian}.  Here,  many subprocesses 
may lead to the final state Lambda so that one has to      
select certain kinematic regions to suppress the unwanted contributions. 
In particular, in order 
to test whether polarized up and down quarks do fragment  into polarized 
Lambdas the rapidity of the produced Lambda has to be large, since at 
high rapidity, $\Lambda$'s are mainly produced through valence 
up and down quarks. 
(We count positive rapidity in the direction of the 
polarized proton beam.)  

The difference of the  
cross sections to produce a Lambda with positive 
helicity through the scattering of a proton with positive/negative 
helicity on an unpolarized proton  is given 
in leading order perturbative QCD (LO pQCD) by
\footnote{Since the relevant spin dependent cross sections 
on the parton level are only 
known in LO we perform a LO calculation here.} 
\begin{eqnarray} 
  E_C\frac{\Delta d\sigma}{d^3p_C} (AB\rightarrow C +X ) 
  & = & 
  E_C\frac{d\sigma}{d^3p_C} (A^\uparrow B\rightarrow C^\uparrow +X ) 
 - 
  E_C\frac{ d\sigma}{d^3p_C} (A^\downarrow B\rightarrow C^\uparrow +X ) 
 \nonumber \\ 
  & = & \sum_{abcd} 
  \int dx_a dx_b dz_c \Delta f_{Aa}(x_a,\mu^2)  
     f_{Bb}(x_b, \mu^2 ) \Delta D_{cC}(z_c, \mu^2) \nonumber \\ 
& &  \frac{\hat{s}}{\pi z^2_c}  \frac{\Delta d\sigma}{d \hat{t}}
 (ab\rightarrow cd ) \,
 \delta (\hat{s} + \hat{t} +\hat{u} ).
\label{cross}   
\end{eqnarray}  
Here, $\Delta f_{Aa}(x_a,\mu^2)$ 
and $f_{Bb}(x_b,\mu^2)$ are the polarized and unpolarized 
distribution functions  of partons $a$ and $b$ in protons $A$ and $B$, 
respectively, at the scale $\mu$. 
$x_a$ and $x_b$ are the corresponding  momentum fractions 
carried by  partons $a$ and $b$. 
$\Delta D_{cC}(z_c,\mu^2 )$ is the polarized fragmentation function of 
parton $c$ into baryon $C$, in our case 
$C =\Lambda$. $z_c$ is the momentum fraction of parton $c$ 
carried by the produced Lambda. $\Delta d\sigma /d \hat{t}$ is the 
difference of the cross sections at the parton level between the 
two processes   $a^\uparrow +b\rightarrow c^\uparrow + d$ and  
$a^\downarrow +b\rightarrow c^\uparrow + d$.     
The unpolarized cross section is given by Eq. (\ref{cross})  
with the $\Delta$'s dropped throughout.  

The Mandelstam variables at the parton level are given 
by  
\begin{equation} 
  \hat{s} =x_a x_b s,  \,\,\,\,
  \hat{t} = - x_a p_\perp \sqrt{s} e^{-y}/z_c, \,\,\,\, 
  \hat{u} =  - x_b p_\perp \sqrt{s} e^{y}/z_c 
\end{equation} 
where, $y$ and $p_\perp$ are  the rapidity and 
transverse momentum of the produced Lambda and $\sqrt{s}$ is the total 
center of mass energy.  
The summation in Eq.(\ref{cross}) runs over all possible 
parton-parton combinations, $qq^\prime \rightarrow qq^\prime$, 
$qg\rightarrow qg$, $q\bar q\rightarrow q\bar q$...  
The elementary unpolarized and polarized cross sections 
can be found in Refs. \cite{Owens2,Stratmann}. 
Performing the   integration in Eq. (\ref{cross}) over 
$z_c$ one obtains 
\begin{eqnarray}  
  E_C\frac{\Delta d\sigma}{d^3p_C} (AB\rightarrow C +X ) 
 & = &  
\sum_{abcd} 
  \int_{x_{amin}}^1 dx_a  
   \int_{x_{bmin}}^1 dx_b   \Delta f_{Aa}(x_a,\mu^2)  
     f_{Bb}(x_b, \mu^2 ) \Delta D_{cC}(z_c, \mu^2) \nonumber \\ 
 & & \frac{1}{\pi z_c} 
 \frac{\Delta d\sigma}{d \hat{t}}
 (ab\rightarrow cd) 
\end{eqnarray} 
with 
\begin{equation} 
    z_c  = \frac{x_\perp}{2x_b} e^{-y} + \frac{x_\perp}{2x_a} e^y, \,\,\,\,\, 
     x_{bmin} = \frac{x_ax_\perp e^{-y}}{2x_a-x_\perp e^y}, \,\,\,\,  
    x_{amin} =  \frac{x_\perp e^y}{2-x_\perp e^{-y}} 
\end{equation} 
where $x_\perp = 2p_\perp /\sqrt{s}$. 

In order to elucidate the kinematics, in Fig.~\ref{fig3},  we plotted 
$z_c$ as a function of $x_a$ and $y$ for two different 
transverse momenta, $p_\perp = 10$ GeV (left) and 
$p_\perp = 30$ GeV (right) and for two different values 
of $x_b$, $x_b=x_{bmin}+0.01$ (top) and  
and $x_b=x_{bmin}+0.1$ (bottom) in Fig.~\ref{fig3}. 
Note, that $z_c$ is maximal both  for $x_b = x_{bmin}$ and 
$x_a=x_{amin}$.  
With increasing rapidity, $y$, both the lower integration limit of 
$x_a$, $x_{amin}$, and the momentum fraction of the fragmenting 
quark transferred to the produced $\Lambda$, $z_c$, increase. 
Hence, large rapidities probe 
the fragmentation of mostly valence quarks into fast 
Lambdas. The dependence on the transverse momenta 
is also shown in Fig.~\ref{fig3}. With increasing $p_\perp$, 
the kinematic boundary is shifted to smaller rapidities and 
the fragmentation of the valence quarks can be studied at lower 
rapidities. This is important since the available 
phase space is limited by the acceptance of the detectors 
at RHIC. However, 
the cross section also decreases with increasing transverse 
momenta, leading to lower statistics.

In Fig.~4, we show the contributions of the various channels 
to the cross section for two different transverse momenta,  
both for inclusive Lambda (4a) and inclusive jet production (4b).  
$gq\rightarrow gq$ stands, for example, for the 
contribution to the cross section coming from 
the subprocess involving a gluon $g$ and a quark $q$ in the 
initial and final states. In $qq\prime \rightarrow  
qq\prime$, the quarks have different flavors and  
$q\bar q\prime \rightarrow   
q\bar q\prime$ is also included. 
Although the kinematics are not exactly the same 
for these two processes 
\footnote{While there is only one integration variable, $x_a$,   
in inclusive jet production, once $p_\perp$ and $y$ are fixed,  
both $x_a$ and $x_b$  
have to be integrated over the allowed kinematic  region 
in inclusive Lambda production, since the produced 
$\Lambda$ carries only a fraction of the parton's 
momentum.} one can 
study the role played by  the fragmentation functions 
by comparing  inclusive Lambda and inclusive jet production.   
In particular, the contributions from channels containing 
two gluons in the final state are suppressed in inclusive 
Lambda production due to the smallness of $D_{g\Lambda}$. 
We note that $D_{g\Lambda}$ has been set to zero at the initial 
scale and is generated  through  evolution. Thus, while 
$qg\rightarrow qg$ is the dominant  channel in inclusive 
Lambda production, both $qg\rightarrow qg$ and 
$gg\rightarrow gg$ are equally important in inclusive 
jet production. There is  some  
ambiguity due to our poor knowledge of the
gluon fragmentation --- larger probabilities 
for $g\rightarrow \Lambda$ will enhance the contributions  
from gluons  in inclusive Lambda production.    
However,  the contribution from the process  
$gg\rightarrow gg$ falls off faster than  
that from $qg\rightarrow qg$ with increasing rapidity,  
since $g(x_a)$ decreases faster than $q(x_a)$ with increasing 
$x_a$ and the Lambdas are produced mainly from valence up and down quarks 
at high rapidities.

Our analysis of the kinematics and the various contributions 
to inclusive Lambda production already indicate that 
Lambda polarization measurements in pp collisions at high rapidities 
are ideally suited to test whether polarized up and down quarks  
may fragment into polarized $\Lambda$'s. 
We calculated the Lambda polarization using our flavor asymmetric  
fragmentation  functions for RHIC energies and for 
$p_\perp = 10$ GeV. Increasing the transverse momentum 
gives similar results, with the only difference that 
$P_\Lambda$ starts to increase at lower rapidities.  
We used the standard set of GRSV LO quark distributions  
for the polarized parton distributions \cite{GRSV} and 
the LO Cteq4 distributions for the unpolarized 
quark distributions \cite{Cteq}.   
The scale, $\mu$, is set equal to $p_\perp$. We  also 
checked that there is only a very weak dependence on the scale 
by calculating the polarization using $\mu=p_\perp/2$ 
and $\mu=2p_\perp$.  The predicted Lambda polarization 
is shown in Fig~\ref{fig5}a. It is positive 
at large rapidities where the contributions of polarized 
up and down quarks dominates the production process.  
At smaller rapidities, where $x_a$ is  small, strange quarks 
also contribute. However, since the ratios of the  
polarized to the unpolarized parton distributions are small 
at small $x_a$  the Lambda polarization is suppressed.  
The result also depends on the parameterization of the polarized 
quark distributions. 
In particular, the polarized gluon distribution 
is not well constrained.  
However, it is  
clear from the kinematics that the ambiguity associated with 
the polarized gluon distributions only effects the results 
at lower rapidities. This can be seen in Fig.~\ref{fig5}b where    
we plot  the contribution  from gluons, up plus down quarks and 
strange quarks to the Lambda polarization.

Next, we contrast 
our prediction with the predictions of various  
$SU(3)$ flavor symmetric models which use  
\begin{equation} 
  D_{u\Lambda}= D_{d\Lambda} = D_{s\Lambda}. 
\label{su3}  
\end{equation}
We fitted the cross sections in $e^+e^-$ annihilation using 
Eq. (\ref{su3})  and the functional form given in Eq. (\ref{para}).   
For the polarized fragmentation functions, we  
discuss two different scenarios:  
The model, $SU(3)_A$ (c.f. Fig. 5a), corresponds to the expectations of the naive 
quark model that only polarized strange quarks can fragment 
into polarized Lambdas 
\begin{equation} 
  \Delta D_{u\Lambda } =   \Delta D_{d\Lambda } =0 \,\,\,\,\,
  \Delta D_{s\Lambda } = D_{s\Lambda}.  
\end{equation} 
It gives essentially zero polarization because the  strange quarks  
contribute at low rapidities where the polarization is 
suppressed.   
Model, $SU(3)_B$ (c.f. Fig. 5a), which was  proposed 
in Ref.\cite{Burk}, is  based on DIS data, and sets 
\begin{equation} 
  \Delta D_{u\Lambda } =   \Delta D_{d\Lambda } = -0.20   
 D_{u\Lambda } \,\,\,\,\,
  \Delta D_{s\Lambda } = 0.60 D_{s\Lambda}.  
\end{equation}  
This model predicts negative Lambda polarization.

Finally, we 
address the problem of Lambdas produced 
through the decay of other hyperons, such as $\Sigma^0$ and $\Sigma^*$. 
In order to estimate the contribution of  
hyperon decays we assume, in the following, that 

(1) the $\Lambda$'s produced through hyperon decay inherit the 
momentum of the parent hyperon   

(2) and that  the total probability to produce 
$\Lambda$, $\Sigma^0$ or  $\Sigma^*$ from 
a certain $uds$ state is given by the $SU(6)$ wave function 
and is independent of the mass of the produced hyperon.   

Further, in order to estimate the polarization transfer 
in the decay process we use the constituent  quark model. 
The polarization can be obtained by noting that the boson  emitted 
in both the  $\Sigma^0\rightarrow \Lambda\gamma$ 
and the $\Sigma^*\rightarrow \Lambda\pi$ decay changes the  
angular momentum of the nonstrange diquark from 
$J=1$ to $J=0$, while  the polarization of the 
spectator strange quark is unchanged. Then, the polarization of the 
$\Lambda$ is determined by the polarization of the 
strange quark in the parent hyperon, since the polarization of the 
$\Lambda$ is exclusively carried by the strange quark in the 
naive quark model. 

First, let us discuss the case when the parent hyperon is 
produced by a strange quark.  
Since the strange quark is always accompanied by a {\it vector} 
$ud$ diquark, in both   $\Sigma^0$  and $\Sigma^*$  
the fragmenation functions of strange quarks into these hyperons 
are  much {\it softer} than  
the corresponding fragmentation function into a $\Lambda$. 
Thus, in the high $z$ limit, the contributions from 
the processes, $s \rightarrow \Sigma^0 \rightarrow \Lambda$ and  
$s \rightarrow \Sigma^* \rightarrow \Lambda$, 
are negligible compared to the direct production, $s \rightarrow \Lambda$. 
Furthermore, both  channels, 
$s \rightarrow \Sigma^0 \rightarrow \Lambda$ and 
$s \rightarrow \Sigma^* \rightarrow \Lambda$,  
enhance the already positive polarization from the direct 
channel, $s \rightarrow \Lambda$.

This is different in the case when the 
parent hyperon is produced by 
an up or down quark.  Both $\Lambda$ and $\Sigma^0$ can be produced 
by an up (down) quark and a  {\it scalar} $ds$ ($us$) 
diquark  --- a process which dominates  in the large $z$ limit. 
(The component with a vector diquark can be neglected in this limit).  
Furthermore,   
the up and down fragmentation function of the $\Sigma^*$ 
are as important as  those of the $\Lambda$ and $\Sigma^0$ 
in the large $z$ limit. This is because  
the  $u$ fragmentation function of $\Sigma^*$ peaks  at about 
$1385/(1010+1385)\approx 0.58$ which is almost the 
same as the peak of the 
{\it scalar} components of the $\Lambda$ and $\Sigma$, which are  
$1115/(890+1115)\approx 0.57$  
and $1190/(890+1190)\approx 0.57$, respectively. 
Thus, for the up and down quark fragmentation, it is important to 
include the $\Lambda$'s from these decay processes.

The relevant probabilities to produce a $\Lambda$ 
with positive and negative polarization from a fragmenting up 
quark with positive polarization and an $ds$ diquark 
are shown in Table \ref{table:3}. We assumed that 
all spin states of the $ds$ diquark are produced with equal 
probabilities. 
The final weights which are 
relevant in the large $z$ limit are set in bold.   
We find that if we include all channels,  
which survive in the large $z$ limit,   
the polarization of the $\Lambda$   
is reduced by a factor of $10/27$ compared to the case where 
only the directly produced $\Lambda$'s are included. 
Since the $\Sigma^*$ decay is a strong decay it is sometimes 
included in the fragmentation function of the 
$\Lambda$. Including only $\Sigma^*$, 
the suppression factor we obtain is   $49/81$.  
(Note that our model predicts that 
$u^\uparrow  (ds)_{0,0}\rightarrow \Lambda$,   
$u^\uparrow (ds)_{0,0}\rightarrow \Sigma^0$  and 
$u^\uparrow (ds)_{1,0} \rightarrow \Sigma^*$  have 
approximately the same $z$ dependence and  are 
approximately equal (up to the Clebsch-Gordon factors) 
since the ratios, $M/(M+M_n)$, have roughly the same 
numerical values.    
Thus, the effect of the $\Sigma^0$ and $\Sigma^*$  decays 
can be taken into account by  a multiplicative  factor.) 
 
In order to illustrate the effect of these decays 
on the final $\Lambda$ polarization, we multiplied 
our results  
with these factors. The results are shown in 
Fig.~6b as dotted lines.  
We note that our implementation of this correction relies on the 
assumptions that the produced $\Lambda$ 
carries all the momentum of the parent hyperon and 
that all states are produced with equal probabilities. 
Since neither of these assumptions is   
strictly valid, we tend to overestimate the importance of  
hyperon decays. Note also that the inclusion of 
$\Sigma^0$ decay in the $SU(3)$ symmetric models 
makes the resulting polarization more negative. 
As a result, even if effects of $\Sigma$ decays are included, large 
discrepancies still persist between our predictions and those 
of $SU(3)$ symmetric models.

\section{Conclusions}  
Measurements of the Lambda polarization at RHIC  
would provide a clear  answer to the question 
of  whether polarized up and down 
quarks can transfer polarization to the final state 
$\Lambda$. We predict {\it positive} 
Lambda polarization at high rapidities,  
in contrast with models 
based on $SU(3)$ flavor symmetry and DIS which predict {\it zero} 
or {\it negative} Lambda polarization.  
Our prediction is  
based on the same physics which led to harder 
up than down quark distributions in the proton and to the 
$\Delta$-N and $\Sigma$-$\Lambda$ mass 
splittings.   
We also estimated the importance of $\Sigma^0$ and 
$\Sigma^*$ decays which tend to reduce the predicted 
$\Lambda$ polarization.

\acknowledgments

This work was partly supported
by the Australian Research Council.  One of the authors [JTL]
was supported in part by National Science Foundation research
contract PHY-9722706. One author [CB] wishes to thank  
the  Indiana University Nuclear Theory Center for its 
hospitality during the time part of this work was
carried out.

\begin{table} 
\caption{Fit parameters  obtained by fitting the $e^+e^-$ data. 
We also parametrized  
$D_{q\Lambda}-D_{\bar q \Lambda}$ and 
$\Delta D_{q\Lambda}-\Delta D_{\bar q \Lambda}$, calculated in the bag.} 
\label{table} 
\begin{tabular}{|l|l|l|l|l|l|} 
\rule[-0.4cm]{0cm}{1cm}
Parameter & $D_{s\Lambda}-D_{\bar s \Lambda}$ 
& $D_{u\Lambda}-D_{\bar u \Lambda}$ 
 & $D_{\bar q\Lambda}$  & 
$\Delta D_{s\Lambda}- \Delta D_{\bar s\Lambda}$ 
& $\Delta D_{u\Lambda}- \Delta D_{\bar u\Lambda}$ 
\rule[-0.4cm]{0cm}{1cm}\\ 
\hline 
\rule[-0.4cm]{0cm}{1cm}
$N$ & $5.81\times 10^9$ & $1.60\times 10^{17}$ & $99.76$ & $3.73\times 
10^{18}$ & $-6.25\times 10^{10}$
\rule[-0.4cm]{0cm}{1cm}\\
\hline 
\rule[-0.4cm]{0cm}{1cm}
$\alpha$ & $21.55$ & $30.49$ & $1.25$ & $21.21$ & $32.48$ 
\rule[-0.4cm]{0cm}{1cm}\\
\hline 
\rule[-0.4cm]{0cm}{1cm}
$\beta$ & $13.60$ & $28.34$ & $11.60$ & $13.38$ & $27.72$ 
\rule[-0.4cm]{0cm}{1cm} \\
\hline 
\rule[-0.4cm]{0cm}{1cm}
$\gamma$ & --- & --- & --- & --- & $0.52$ 
\rule[-0.4cm]{0cm}{1cm} \\
\hline 
\end{tabular} 
\end{table} 

\begin{table}
\caption{Fit parameters obtained by fitting the $e^+e^-$ data and 
asumming that the fragmentation fucntions are flavor symmetric.}
\begin{tabular}{|l|l|l|} 
\rule[-0.4cm]{0cm}{1cm}
Parameter & $D_{q\Lambda}-D_{\bar q \Lambda}$ 
 & $D_{\bar q\Lambda}$   
\rule[-0.4cm]{0cm}{1cm}\\ 
\hline 
\rule[-0.4cm]{0cm}{1cm}
$N$ & $1.92\times 10^4$ & $99.76$ 
\rule[-0.4cm]{0cm}{1cm}\\
\hline 
\rule[-0.4cm]{0cm}{1cm}
$\alpha$ & $7.47$ & $1.25$ 
\rule[-0.4cm]{0cm}{1cm}\\
\hline 
\rule[-0.4cm]{0cm}{1cm}
$\beta$ & $8.06$ & $11.60$ 
\rule[-0.4cm]{0cm}{1cm} \\
\end{tabular} 
\end{table}

\begin{table*}
\let\tabularsize\footnotesize

\caption{Different channels for the
production of  $\Lambda$ hyperons from a positively polarized
up quark and a $ds$ diquark. It is  assumed that
all spin states of the $ds$ diquark are produced with the
{\it same} probabilities. $\Sigma^{*\uparrow}$ and
$\Sigma^{*\Uparrow}$  stand for the $1/2$ and $3/2$ spin
component of the $\Sigma^*$. See text for further details.
}
\label{table:3}
\centering
\begin{tabular}[t]{||l||c|c||c|c||c|c||c|c||c|c||c|c||c|c||c|c||c|c||c|c||c|c||
c|c||}
\hline \hline
\rule[-0.4cm]{0cm}{1cm}
$u(ds)$ states &
 \multicolumn{6}{c||}{$u^\uparrow (ds)_{0,0}$} &
 \multicolumn{6}{c||}{$u^\uparrow (ds)_{1,1}$} &
 \multicolumn{6}{c||}{$u^\uparrow (ds)_{1,0}$} &
 \multicolumn{6}{c||}{$u^\uparrow (ds)_{1,-1}$} \\ \hline
\rule[-0.4cm]{0cm}{1cm}
relative weights &
 \multicolumn{6}{c||}{$\frac{1}{4}$} &
 \multicolumn{6}{c||}{$\frac{1}{4}$} &
 \multicolumn{6}{c||}{$\frac{1}{4}$} &
 \multicolumn{6}{c||}{$\frac{1}{4}$} \\ \hline
\rule[-0.4cm]{0cm}{1cm}
products & \multicolumn{2}{c||}
  {$\Lambda ^\uparrow$}    &
\multicolumn{2}{c||} {$\Sigma^{0\uparrow }$}   &
\multicolumn{2}{c||} {$\Sigma^{*0\uparrow }$}  &
\multicolumn{6}{c||} {\mbox{\hspace{0.5cm}}
  $\Sigma^{*0\Uparrow }$ \mbox{\hspace{0.5cm}} } &
\multicolumn{2}{c||} {$\Lambda ^\uparrow $}    &
\multicolumn{2}{c||} {$\Sigma^{0\uparrow }$}   &
\multicolumn{2}{c||} {$\Sigma^{*0\uparrow }$}  &
\multicolumn{2}{c||} {$\Lambda ^\downarrow $}  &
\multicolumn{2}{c||} {$\Sigma^{0\downarrow }$}   &
\multicolumn{2}{c||} {$\Sigma^{*0\downarrow }$}  \\ \hline
\rule[-0.4cm]{0cm}{1cm}
 relative weights
  &\multicolumn{2}{c||}{$\frac{1}{4}$} &
\multicolumn{2}{c||}{$\frac{3}{4}$} &
\multicolumn{2}{c||}{0}
&\multicolumn{6}{c||}{1}
  &
\multicolumn{2}{c||}{$\frac{1}{4}$}&
\multicolumn{2}{c||}{$\frac{1}{12}$}&
\multicolumn{2}{c||}{$\frac{2}{3}$}&
\multicolumn{2}{c||}{$\frac{1}{2}$}&
\multicolumn{2}{c||}{$\frac{1}{6}$}&
\multicolumn{2}{c||}{$\frac{1}{3}$} \\ \hline
\rule[-0.4cm]{0cm}{1cm}
decay products &
\multicolumn{2}{c||}{$\Lambda^\uparrow$}&
$\Lambda^\uparrow$ & $\Lambda^\downarrow$ &
\multicolumn{2}{c||}{$-$}&
\multicolumn{3}{c|}{$\Lambda^\uparrow$} &
\multicolumn{3}{c||}{$\Lambda^\downarrow$}&
$\Lambda^\uparrow$ & $\Lambda^\downarrow$ &
$\Lambda^\uparrow$ & $\Lambda^\downarrow$ &
$\Lambda^\uparrow$ & $\Lambda^\downarrow$ &
$\Lambda^\uparrow$ & $\Lambda^\downarrow$ &
$\Lambda^\uparrow$ & $\Lambda^\downarrow$ &
$\Lambda^\uparrow$ &
\multicolumn{1}{c||}{$\Lambda^\downarrow$} \\ \hline
\rule[-0.4cm]{0cm}{1cm}
relative weights & \multicolumn{2}{c||}{1} &
 $\frac{1}{3}$ & $\frac{2}{3}$ & \multicolumn{2}{c||}{$0$} &
\multicolumn{3}{c|}{\phantom{xx}1\phantom{xx}} &
\multicolumn{3}{c||}{0} &
1 & 0 & $\frac{1}{3}$ & $\frac{2}{3}$ & $\frac{2}{3}$ & $\frac{1}{3}$
& 0 & 1 & $\frac{2}{3}$ & $\frac{1}{3}$ & $\frac{1}{3}$ & $\frac{2}{3}$
\\ \hline
\rule[-0.4cm]{0cm}{1cm}
final weights & \multicolumn{2}{c||}{${\bf \frac{1}{16}}$} &
 ${\bf \frac{1}{16}}$ & ${\bf \frac{1}{8}}$ & \multicolumn{2}{c||}{0} &
\multicolumn{3}{c|}{${\bf \frac{1}{4}}$} &
\multicolumn{3}{c||}{0} &
$\frac{1}{16}$ & 0 & $\frac{1}{144}$ & $\frac{1}{72}$ &
${\bf \frac{1}{9}}$ & ${\bf \frac{1}{18}}$
& 0 & $\frac{1}{8}$ & $\frac{1}{36}$ &
$\frac{1}{72}$ & ${\bf \frac{1}{36}}$ &
${\bf \frac{1}{18}}$ \\ \hline
%=====================================================================
\end{tabular}
\end{table*}

\begin{figure}
\psfig{figure=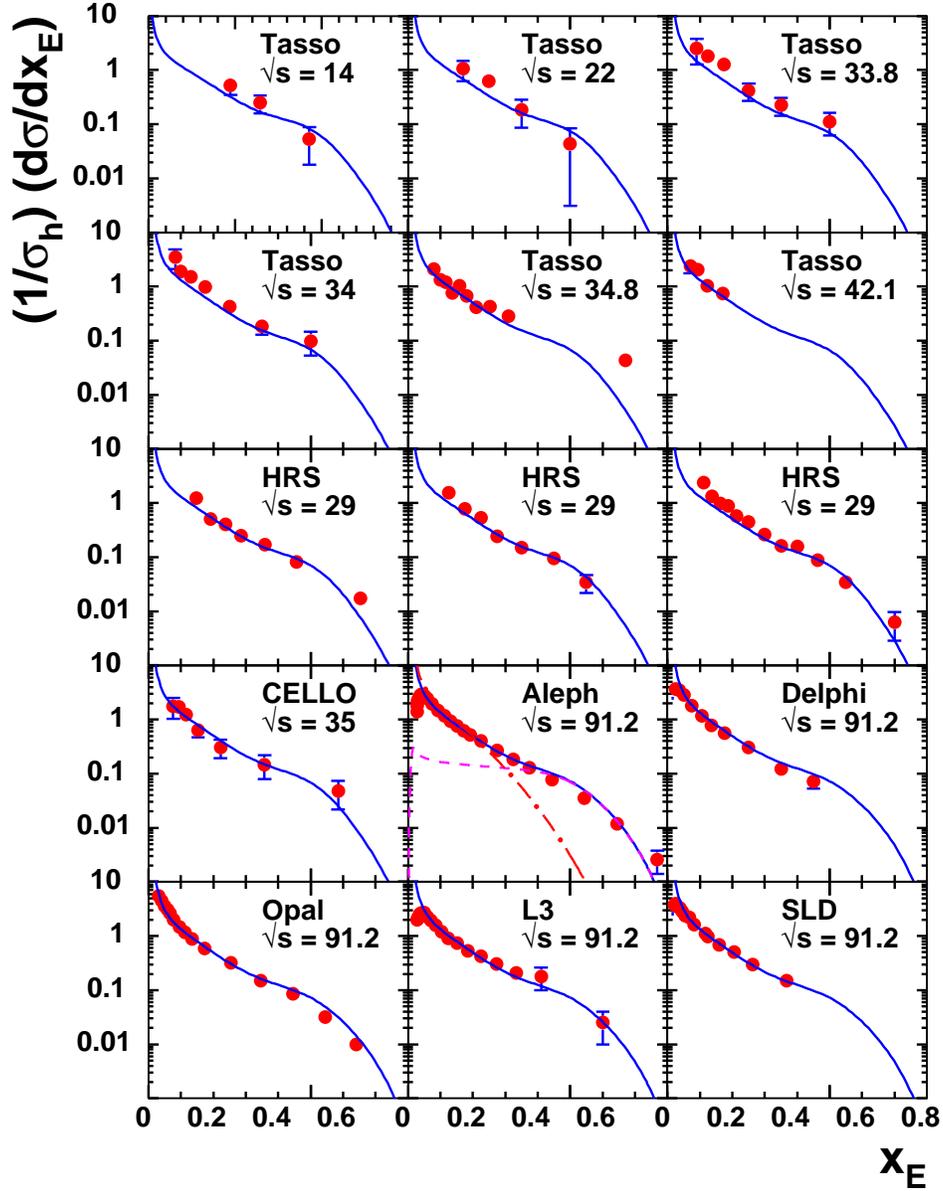,height=18.cm}
\caption{Inclusive Lambda production in $e^+e^-$ annihilation. 
The solid lines are the result of the global fit. They contain two parts, 
the fixed contributions from  $D_{q\Lambda} - D_{\bar q\Lambda}$ 
calculated in the bag (dashed line only shown for the Aleph data) 
and $D_{\bar q\Lambda}$ obtained from the fit (dash-dotted line). 
$x_E$ is defined as $x_E=2E_\Lambda/\sqrt{s}$ where 
$E_\Lambda$ is  the energy 
of the produced Lambda in the $e^+e^-$ center of mass frame  
and $\sqrt{s}$ is the total center of mass enery.   }
\label{fig1}
\end{figure}

\begin{figure}
\psfig{figure=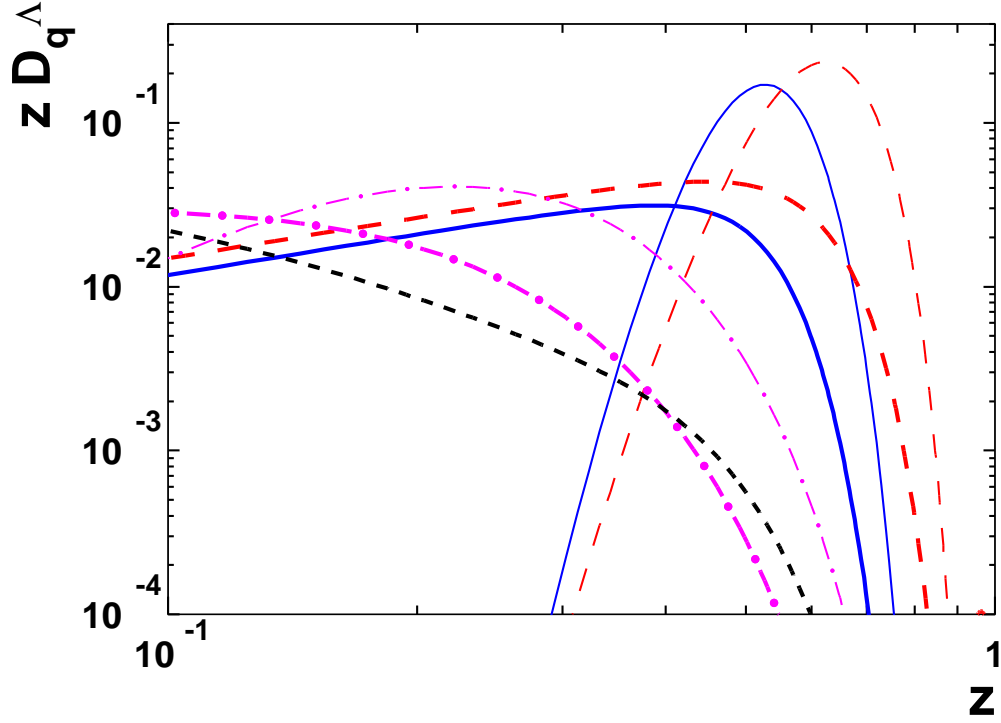,height=11.cm}
\caption{Fragmentation functions. The solid and dashed lines stand for the 
calculated fragmentation  functions of up and strange quarks  into 
Lambda baryons through production of a Lambda and an anti-diquark 
and correspond to $D_{u\Lambda}- D_{\bar u \Lambda}$ 
and  $D_{s\Lambda}- D_{\bar s \Lambda}$, respectively. 
The dash-dotted line represents the contributions from higher 
intermediate states, and is obtained by fitting the 
$e^+e^-$ data and corresponds to $D_{\bar q\Lambda}$.   
The short dashed line is the gluon fragmentation function.  
The light and heavy lines are the fragmentation functions at 
the scales $Q^2=\mu^2$ and $Q^2=M_Z^2$, respectively. 
Note that $D_{g\Lambda}=0$ at $Q^2=\mu^2$.} 
\label{fig2}
\end{figure}

\begin{figure}
\psfig{figure=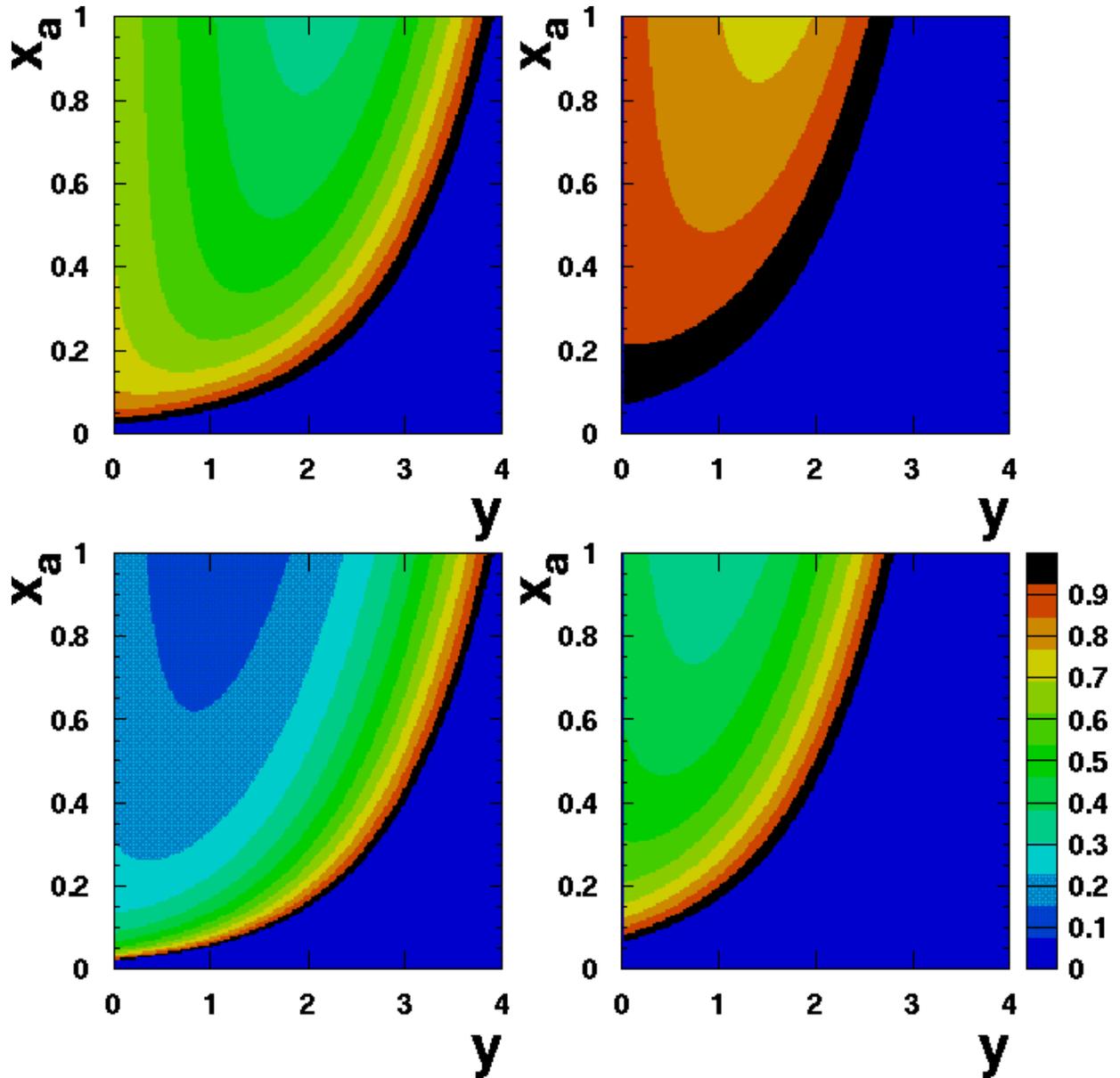,height=16.cm}
\caption{$z_c$ as a function of $x_a$ and $y$ for two different 
transverse momenta, $p_\perp = 10$ GeV (left) and 
$p_\perp = 30$ GeV (right) and for two different values 
of $x_b$, $x_b=x_{bmin}+0.01$ (top) and  
and $x_b=x_{bmin}+0.1$ (bottom). } 
\label{fig3}
\end{figure}

\begin{figure}
\psfig{figure=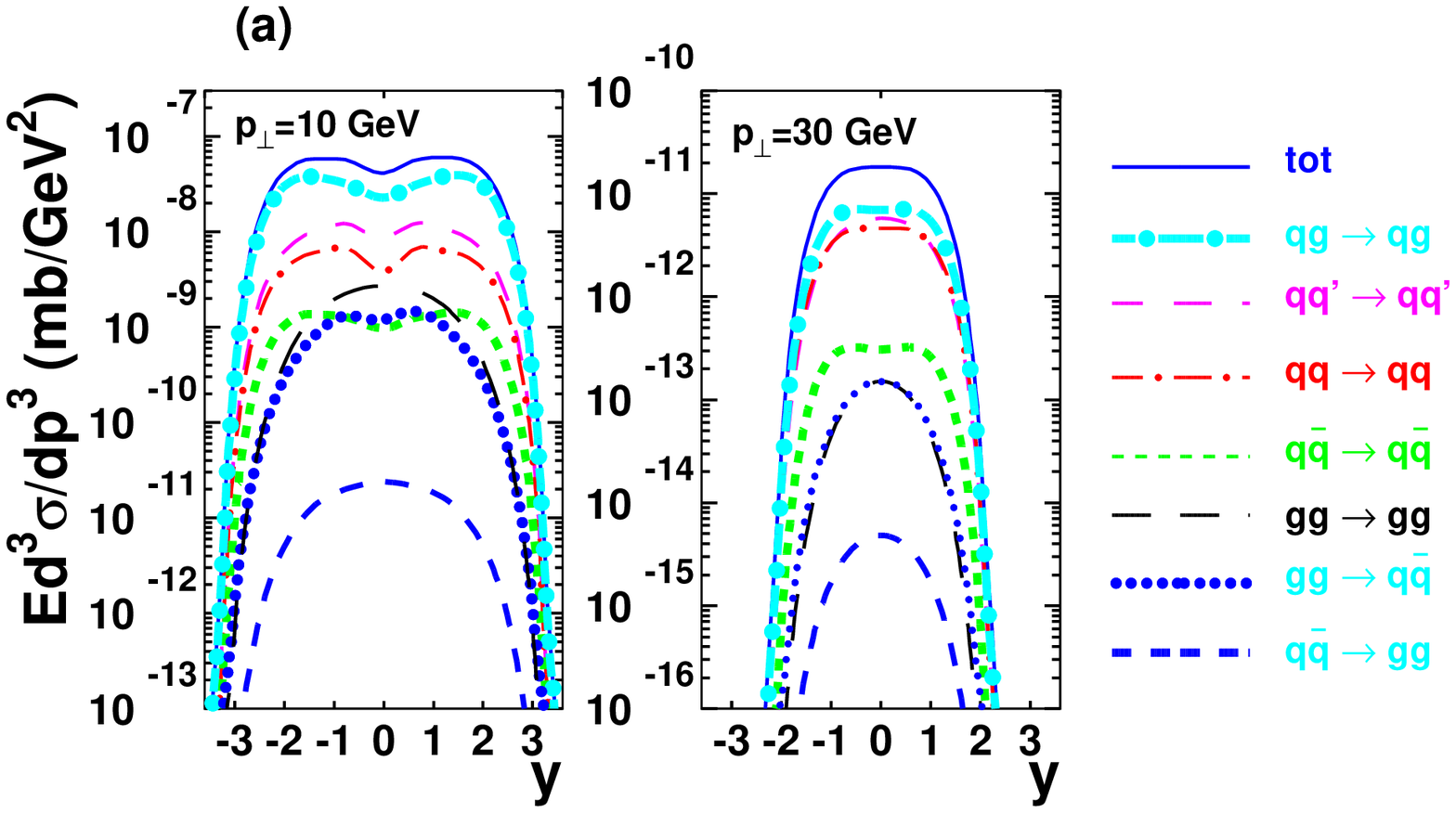,height=10.cm}

\psfig{figure=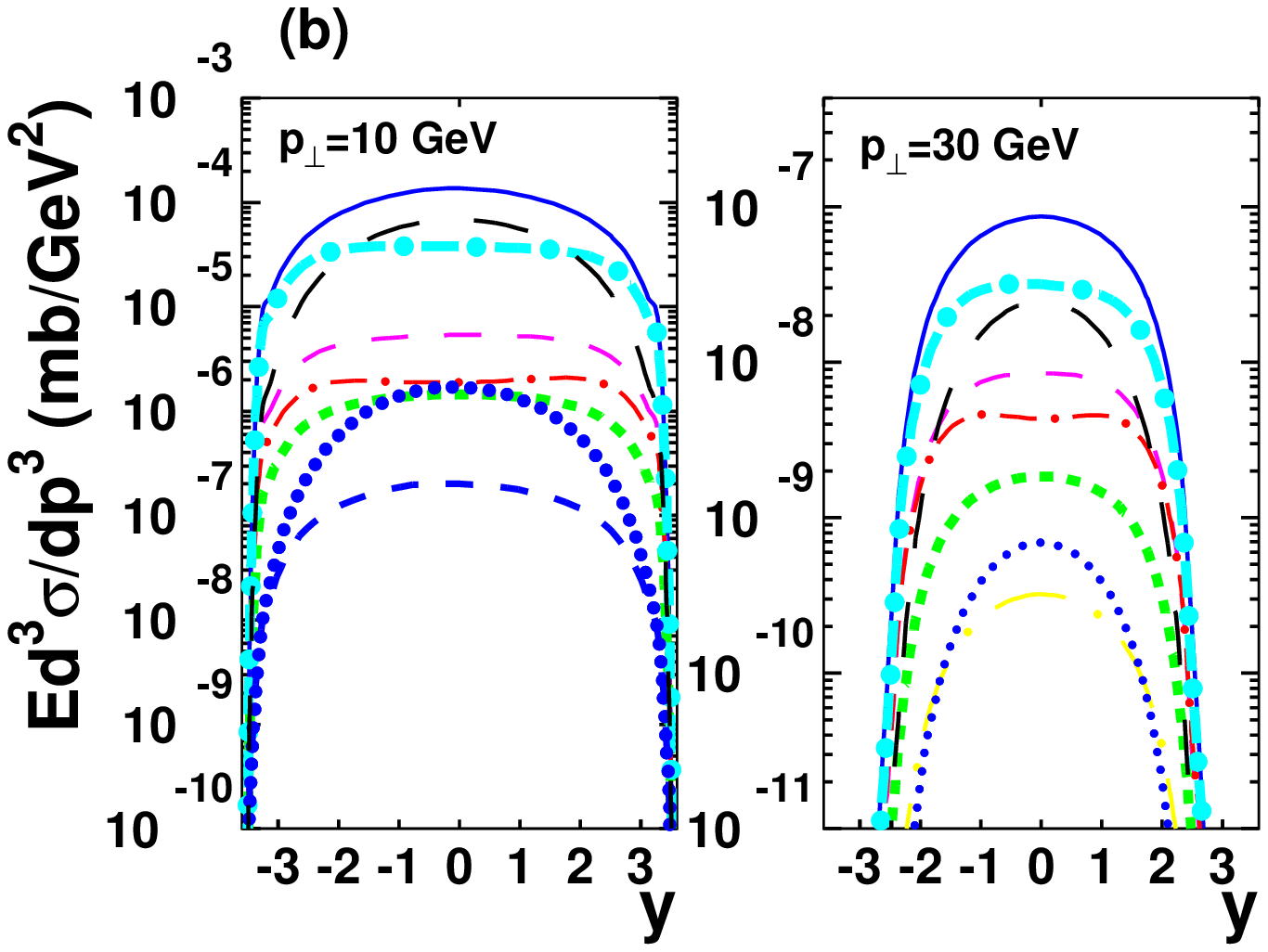,height=10.cm}
\caption{Contributions from the various channels (a)  to the  
inclusive Lambda production 
cross section ($pp\rightarrow \Lambda +X$) 
and (b) to the inclusive jet production 
cross section ($pp\rightarrow jet +X$) 
at $p_\perp =10$ GeV (left) and 
$p_\perp =30$ GeV (right) at $\sqrt{s}=500$ GeV. } 
\label{fig4a}
\end{figure}

\begin{figure}
\psfig{figure=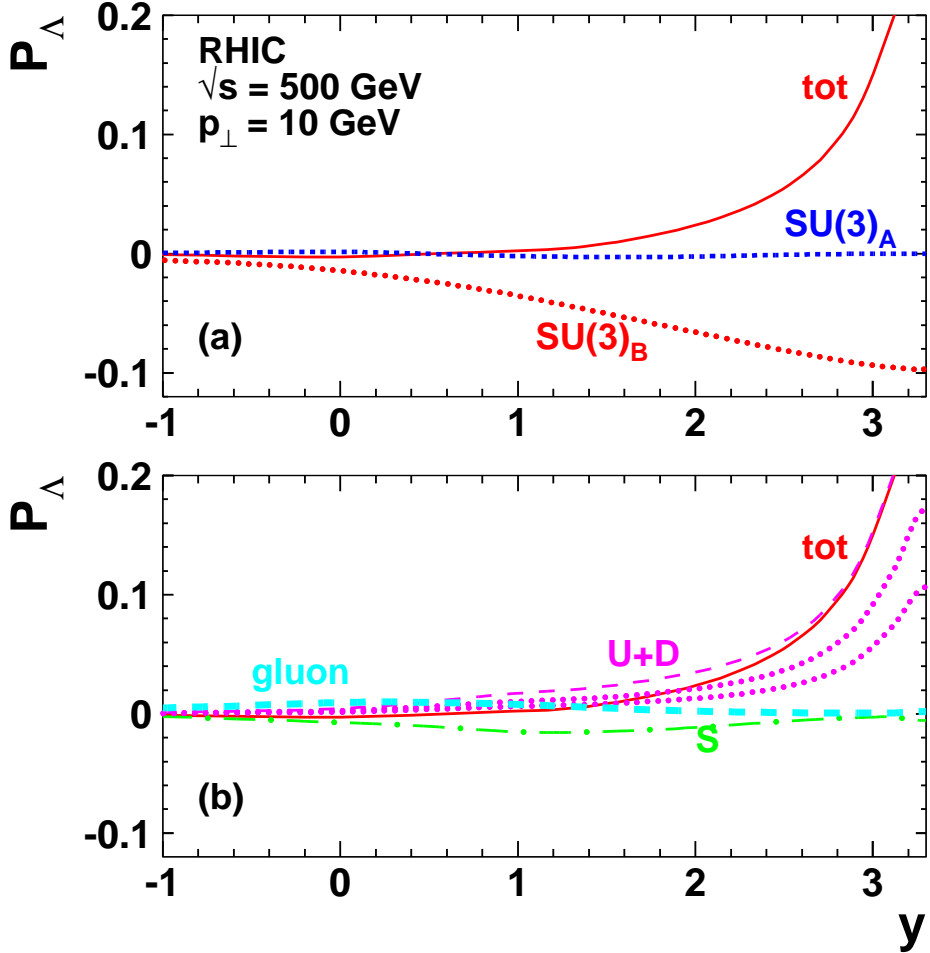,height=14.cm}
\caption{Lambda polarization at RHIC. 
(a) The solid line represents our prediction.  
The predictions  of $SU(3)$ symmetric fragmentation models
are  shown for comparision. 
The model labeled as $SU(3)_A$ is based on the
quark model expectation that only 
the polarized strange quark may fragment into
polarized  Lambdas, while $SU(3)_B$, is based on
DIS data. 
(b) Contributions of different flavors to the 
$\Lambda$-polarization. The light dashed, dash-dotted 
and heavy dashed lines 
stand for  the contributions from up plus down, from
strange  and from gluon fragmentation, respectively, 
as calculated here.  
The estimated polarization including 
both $\Sigma^0$ and $\Sigma^*$ (lower dotted line) and 
only $\Sigma^*$ (upper dotted line) decays are also shown.  
See text for further details. 
}
\label{fig5}
\end{figure}

\end{document}